\documentclass[12pt,preprint]{aastex}

\slugcomment{Accepted for publication in ApJ}

\shorttitle{YSOVAR: ONC mid-IR variability}
\shortauthors{Morales-Calder\'on et al.}

\begin{document}


\title{YSOVAR: the first sensitive, wide-area, mid-IR photometric monitoring of the ONC}



\author{M. Morales-Calder\'on\altaffilmark{1}, J. R. Stauffer\altaffilmark{1}, L. A. Hillenbrand\altaffilmark{2}, R. Gutermuth\altaffilmark{3,4}, I.~Song\altaffilmark{5}, L.~M.~Rebull\altaffilmark{1}, P.~Plavchan\altaffilmark{6}, J.~M.~Carpenter\altaffilmark{2}, B. A. Whitney\altaffilmark{7}, K. Covey\altaffilmark{8,9,10}, C. Alves de Oliveira\altaffilmark{11}, E. Winston\altaffilmark{12}, M. J. McCaughrean\altaffilmark{12,13}, J. Bouvier\altaffilmark{33}, S. Guieu\altaffilmark{14}, F.~J.~Vrba\altaffilmark{15}, J.~Holtzman\altaffilmark{16}, F. Marchis\altaffilmark{17,18}, J. L. Hora\altaffilmark{19}, L. H. Wasserman\altaffilmark{20}, S.~Terebey\altaffilmark{21}, T.~Megeath\altaffilmark{22}, E.~Guinan\altaffilmark{23}, J. Forbrich\altaffilmark{19}, N. Hu\'elamo\altaffilmark{24}, P.~Riviere-Marichalar\altaffilmark{24}, D.~Barrado\altaffilmark{24,25}, K.~Stapelfeldt\altaffilmark{26}, J. Hern\'andez\altaffilmark{27}, L.~E.~Allen\altaffilmark{28}, D. R. Ardila\altaffilmark{29}, A. Bayo\altaffilmark{14}, F. Favata\altaffilmark{13}, D. James\altaffilmark{30,31}, M.~Werner\altaffilmark{26}, and K. Wood\altaffilmark{32}}


\email{mariamc@ipac.caltech.edu}


\altaffiltext{1}{Spitzer Science Center, California Institute of Technology, Pasadena, CA 91125}
\altaffiltext{2}{Astronomy Department, California Institute of Technology, Pasadena, CA 91125}
\altaffiltext{3}{Five College Astronomy Dept., Smith College, Northampton, MA  01063}
\altaffiltext{4}{Dept. of Astronomy, University of Massachusetts, Amherst, MA  01003}
\altaffiltext{5}{Department of Physics and Astronomy, University of Georgia, Athens, GA 30602}
\altaffiltext{6}{NASA Exoplanet Science Institute, California Institute of Technology, Pasadena, CA 91125}
\altaffiltext{7}{Space Science Institute, 4750 Walnut St. Suite 205, Boulder, CO 80301}
\altaffiltext{8}{Cornell University, Department of Astronomy, 226 Space Sciences Building, Ithaca, NY 14853, USA}
\altaffiltext{9}{Hubble Fellow}
\altaffiltext{10}{Visiting Researcher, Department of Astronomy, Boston University, Boston, MA 02215, USA}
\altaffiltext{11}{Laboratoire d'Astrophysique de Grenoble, Observatoire de Grenoble, 38041 Grenoble Cedex 9, France.}
\altaffiltext{12}{Astrophysics Group, College of Engineering, Mathematics, and Physical Sciences, U. of Exeter, UK}
\altaffiltext{13}{European Space Agency, Research \& Scientific Support Dept., ESTEC, Noordwijk, The Netherlands.}
\altaffiltext{14}{European Southern Observatory, Alonso de C\'ordova 3107, Vitacura - Santiago, Chile. }
\altaffiltext{15}{U.S. Naval Observatory, Flagstaff Station, 10391 W. Naval Observatory Road, Flagstaff, AZ 86001-8521.}
\altaffiltext{16}{New Mexico State University, Box 30001/ MSC 4500,  Las Cruces NM 88003.}
\altaffiltext{17}{UC Berkeley, Department of Astronomy, 601 Campbell Hall, Berkeley CA 94720 US}
\altaffiltext{18}{SETI Institute, Carl Sagan Center, 189 N San Bernado Av, Mountain View CA 94043 US}
\altaffiltext{19}{Havard-Smithsonian Center of Astrophysics, 60 Garden Street, Cambridge, MA 02138.}
\altaffiltext{20}{Lowell Observatory, 1400 West Mars Hill Road, Flagstaff, AZ 86001.}
\altaffiltext{21}{Dept. of Physics and Astronomy, California State University at Los Angeles, Los Angeles, CA 90032.}
\altaffiltext{22}{The University of Toledo, 2801 West Bancroft Street, Toledo, Ohio 43606.}
\altaffiltext{23}{Dept. of Astronomy and Astrophysics, Villanova University, Villanova, 19085, PA, USA.}
\altaffiltext{24}{Centro de Astrobiolog\'ia (INTA-CSIC); LAEFF, P.O. Box 78, E-28691 Villanueva de la Canada, Spain.}
\altaffiltext{25}{Calar Alto Observatory, Centro Astron\'omico Hispano Alem\'an, Almer\'ia, Spain.}
\altaffiltext{26}{Jet Propulsion Laboratory, California Institute of Technology, Pasadena, CA 91109.}
\altaffiltext{27}{Centro de Investigaciones de Astronom\'ia, Apdo. Postal 264, M\'erida 5101-A, Venezuela.}
\altaffiltext{28}{National Optical Astronomy Observatory, 950 North Cherry Avenue, Tucson, AZ 85719, USA.}
\altaffiltext{29}{NASA Herschel Science Center, California Institute of Technology, Pasadena, CA 91125.}
\altaffiltext{30}{Hoku Kea Observatory, Dept. of Physics \& Astronomy, U. of HawaiÔi at Hilo, Hilo, HI 96720, USA}
\altaffiltext{31}{Cerro Tololo Inter-American Observatory, Casilla 603, La Serena, CHILE}
\altaffiltext{32}{School of Physics \& Astronomy, U. of St Andrews, North Haugh, St Andrews, Fife, KY16 9AD, UK}
\altaffiltext{33}{UJF-Grenoble 1 / CNRS-INSU, Institut de PlanŽtologie et d'Astrophysique de Grenoble (IPAG) UMR 5274, Grenoble, F-38041, France }


\begin{abstract}
We present initial results from time series imaging at infrared wavelengths of 0.9 sq. degrees in the Orion Nebula Cluster (ONC). During Fall 2009  we obtained 81 epochs of Spitzer 3.6 and 4.5~$\mu$m data over 40 consecutive days.
We extracted light curves with $\sim$3\% photometric accuracy for $\sim$2000 ONC members ranging from several solar masses down to well below the hydrogen burning mass limit.  For many of the stars, we also have  time-series photometry obtained at optical ($I_c$) and/or near-infrared (JK$_s$) wavelengths. Our data set can be mined to determine stellar rotation periods, identify new pre-main-sequence (PMS) eclipsing binaries, search for new substellar Orion members, and help better determine the frequency of circumstellar disks as a function of stellar mass in the ONC.  Our primary focus is the unique ability of 3.6 \& 4.5 $\mu$m variability information   to improve our understanding of inner disk processes and structure in the Class I and II young stellar objects (YSOs).
In this paper, we provide a brief overview of the YSOVAR Orion data obtained in Fall 2009, and
we highlight our light curves for AA-Tau analogs -- YSOs with narrow dips in flux, most probably due to disk density structures passing through our line of sight. Detailed follow-up observations are needed in order to better quantify the nature of the obscuring bodies and what this implies for the structure of the inner disks of YSOs.

\end{abstract}





\section{Introduction}

The Orion star forming region has been the subject of more low mass young star variability studies than any other region.  While \citet{Haro69}, \citet{Herbig72}, and \citet{Walker78} all conducted pioneering studies of the variability of young stars in Orion, the modern era using two-dimensional imaging cameras was initiated by Herbst and his team \citep{Attridge92,Choi96}.  They obtained multi-year, optical time series photometry of several regions of the ONC, eventually obtaining light curves for hundreds of PMS stars, often spanning several years. Those data were primarily used to derive rotation periods -- made possible by the fact that the photospheres of young stars are often heavily spotted (cold and/or hot spots), resulting in flux variations modulated at the star's rotation period.  Many other groups subsequently obtained time series photometry of other portions of the ONC, using optical \citep{Stassun99, Rebull01, Herbst02Orion, Irwin07} and near-IR \citep{Carpenter01} imaging data. In many cases, the light curve shapes are well fit by hot or cold spots, normally at moderately high latitudes since the light curves are seldom ``flat-bottomed." However, for a significant fraction of the light curves, particularly those in the near-IR, 
spots do not seem to provide a good explanation of the observed variability \citep{Carpenter01}.

Time series photometry of YSOs can also address issues other than rotational velocities such as flares and flare frequency \citep{Parsamian93}, disk instabilities \citep{Fedele07,Herbst10,Plavchan08}, and eclipsing binaries as tests of PMS isochrones \citep{Stassun04,Stassun06,Cargile08}.  
Mid-IR time series photometry of YSOs could, in principle, detect variations in the temperature or spatial distribution of warm dust in the inner circumstellar disks of these stars, thereby informing models of disk evolution and perhaps of planet formation. We recently completed the first sensitive, wide-area, mid-IR
time-series photometric monitoring program for the ONC using IRAC \citep{Fazio04} on the Spitzer Space Telescope \citep{Werner04}.  We report here some early results from that program, with more detailed analysis of specific classes of variability to follow.

\section{Observations}


\subsection{Spitzer Observations}

Our GO6 Exploration Science program YSOVAR (Young Stellar Object VARiability) provides the first large-scale survey of mid-IR photometric variability of YSOs. In Fall 2009, we used $\sim$250 hours of warm Spitzer observing time to monitor $\sim$0.9 deg$^2$ of the ONC at 3.6 and 4.5 $\mu$m. 

The observed area was broken into five segments with a central region of $\sim$20$\arcmin\times$25$\arcmin$ and four flanking fields. The central part was observed in full array mode with 1.2 seconds exposure time and 20 dithering positions to avoid saturation by the bright nebulosity around the Trapezium stars. The remaining four segments of the map were observed in High Dynamic Range mode with exposure times of 10.4 and 0.4 seconds, and 4 dithering positions. A summary of the observations appears in Table~\ref{Data}.

These observations were taken for 40 days in Fall 2009, with $\sim$2 epochs each day 
but with an interval between observations that varied both by design (to reduce period-aliasing problems) and due to constraints imposed by other Spitzer programs that were executed during the same campaigns. We used the IDL package Cluster Grinder \citep{Gutermuth09} which, starting from the BCD images released by the Spitzer Science Center, builds the combined mosaic for each epoch and performs aperture photometry on the mosaics. In the present work the BCD images used correspond to the S18.12 software build (the zero point magnitudes are 19.30 and 18.67 for 1 DN/s total flux at 3.6 and 4.5 $\mu$m, respectively). Future versions of the pipeline will include improvements, most importantly using better linearity corrections. 

\subsection{Ground-based Data}

In order to complement our Spitzer data, we also obtained $I_c$, $J$, and $K_s$ contemporaneous photometry, usually for smaller areas within the Spitzer mosaic. The main datasets come from the UKIRT Wide Field CAMera (WFCAM), the CFHT Wide field InfraRed Camera (WIRCam), the Steward Observatory Super-LOTIS (Livermore Optical Transient Imaging System) robotic telescope, and the NMSU/APO 1m telescope.  A summary of the principal characteristics 
of each dataset appears in Table~\ref{Data}. A deeper analysis on data reduction of this data set plus additional data collected for this project will be described in Gutermuth et al. (2011, in prep.).

We performed differential aperture photometry on the ground-based data. In each dataset, for each object, an artificial reference level was created by selecting 30 nearby isolated stars; we iteratively eliminated those with larger photometric scatter  or evidence of variability. Finally, the time series are incorporated into a database that will become publicly available in early 2011 (http://ysovar.ipac.caltech.edu/). This database includes: a color image of the surveyed field of view, fits files for the 3.6 and 4.5 $\mu$m mosaics from a single epoch, a table for all ONC stars providing the name, aliases, RA, Dec, J,H,K, [3.6], [4.5], [5.8], [8.0], and [24] information as in Megeath et al. (2011, in prep.), a file with plots of multi-wavelength light curves, and the actual time series for all ONC members either as ascii files or by querying the database. Alternatively, the time series can be found in Table\ref{TimeSeries}.




\section{Variability statistics and types}

Light curves were extracted in both IRAC bands for 1249 Class I and II previously known Orion YSOs with IR excess (Megeath et al. 2011, in prep.) and inferred masses between 0.03 and $\sim$2 M$_\sun$, plus 820 other likely Orion members. This latter set of objects is formed by sources that fulfill at least one of the following conditions: a) has been labeled as a member by proper motion studies such as \citet{Parenago54,Jones88,McNamara76,Tian96}, b) has been labeled as a member according to \citet{H98}, c) has been labeled as a member by X-ray studies \citep{,Getman05}, d) has H$_\alpha$ profiles typical of weak-lined T~Tauris (WTTs) \citep{Tobin09,Furesz08},  e) has been labeled as variable by \citet{Carpenter01}, or f) finally we have included all the sources that have been claimed to be periodic variables.  Most of these stars are WTTs according to the Megeath et al. photometry, though some may be sources with excesses that escaped previous detection. In addition to these 2069 Orion stars, another 147 Class I and Class II sources plus 209 Orion members without IR excess have light curves in just one of the two IRAC bands.

We estimated the noise in the IRAC light curves by measuring the typical scatter for all the objects in the field. In that way, we estimate a relative per channel photometric accuracy better than 3\% down to [3.6]=14 and better than 10\% for objects as faint as [3.6]=16. For the $J$ band, we derive accuracies of $\sim$2\% down to $J$=16.5 and of $\sim$10\% to J=18.5 for the ensemble. Note that the accuracy varies significantly over the surveyed field due to nebular emission.

\begin{figure}
\epsscale{1.}
\plotone{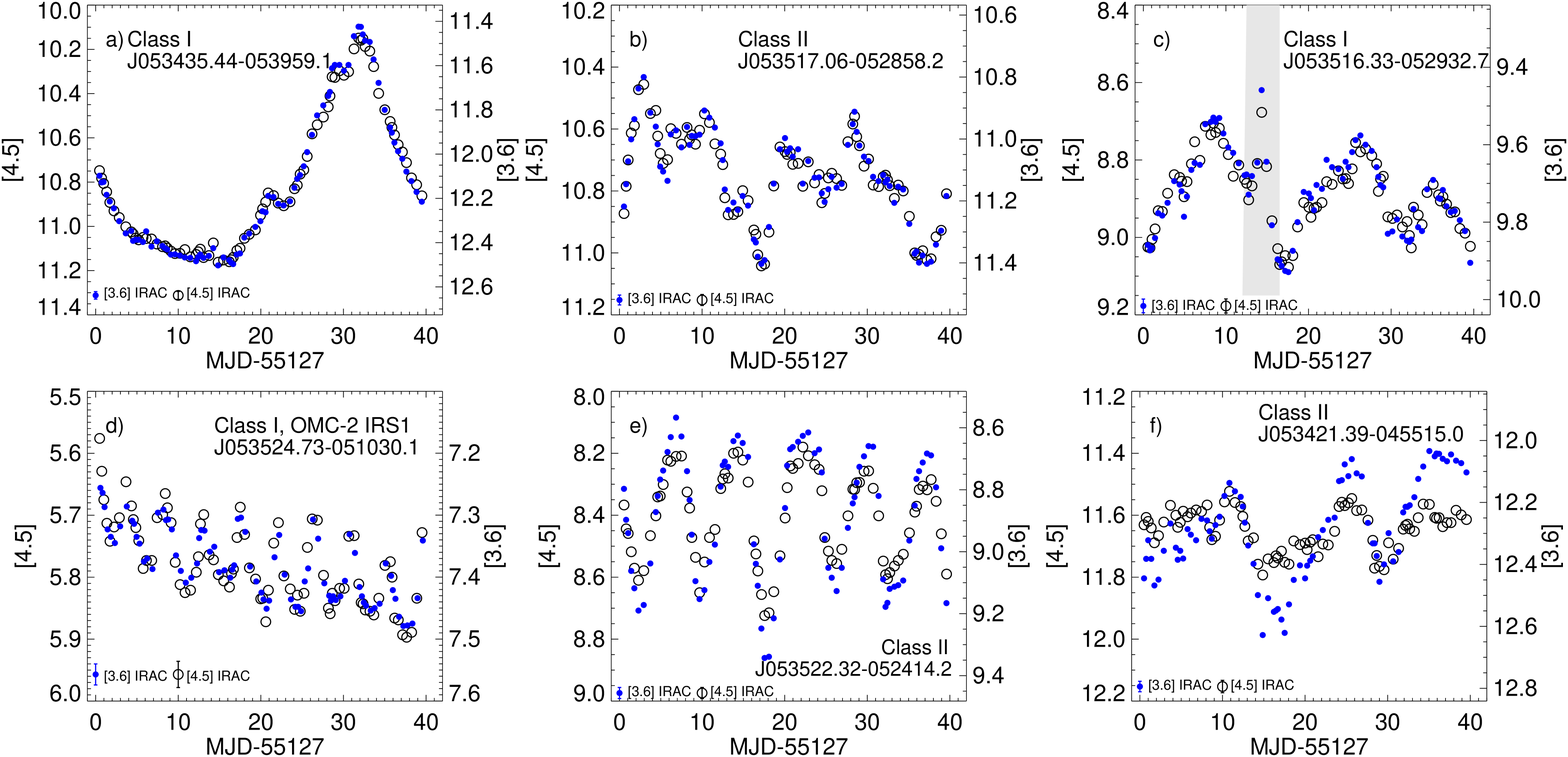}
\caption{Sample IRAC light curves for several stars showing the range of different variability types found in our data: a) slow drifts in brightness that may be caused by a slow change in accretion rate or by self-shadowing in a warped disk; b) rapid chaotic flux variations probably caused by changes in the accretion geometry; c) flares; d) periodic variations mixed with other longer term changes; e) periodic variations arising from photospheric spots and/or disk warps; f) changes in color that may be produced by variable extinction. The symbols are $\bullet$: [3.6]; $\circ$: [4.5]. [3.6] and [4.5] magnitudes have been plotted in the right and left vertical axis respectively.  \label{IRAClcs}}
\end{figure}

Approximately 70\% of the YSOs with IR excess are variable based primarily on the Stetson index calculated using both IRAC channels together (\citet{Welch93}; see Table~\ref{statistics} for statistics). The variable sources were visually inspected and, in addition to the Stetson variables, we considered as variables as well a few YSOs for which a period could be found even when the Stetson index alone would not reflect the variability. About 3\% of the variable YSOs have been selected in this way. Usually these sources have very low amplitude however, most of them have previously detected periods that match the ones derived here and thus we believe the variability is real. The variable stars exhibit diverse behavior with time, including slow drifts in brightness over one month or longer that may be caused by a slow change in accretion rate or by self-shadowing in a warped disk (Fig.~\ref{IRAClcs}a); rapid chaotic flux variations probably caused by changes in the accretion geometry (Fig.~\ref{IRAClcs}b); flares (Fig.~\ref{IRAClcs}c); stars with lightcurves that look ``periodic" for part but not all of the 40 day period or for which the periodic variations are mixed with other longer term changes (Fig.~\ref{IRAClcs}d); periodic or pseudo-periodic variations arising from photospheric spots and/or disk warps (Fig.~\ref{IRAClcs}e); changes in color from variable extinction (?) (Fig.~\ref{IRAClcs}f); and stars that show narrow or broad dips in the observed flux superimposed on an otherwise nearly constant flux level (see next section). 

\begin{figure}
\epsscale{1.}
\plotone{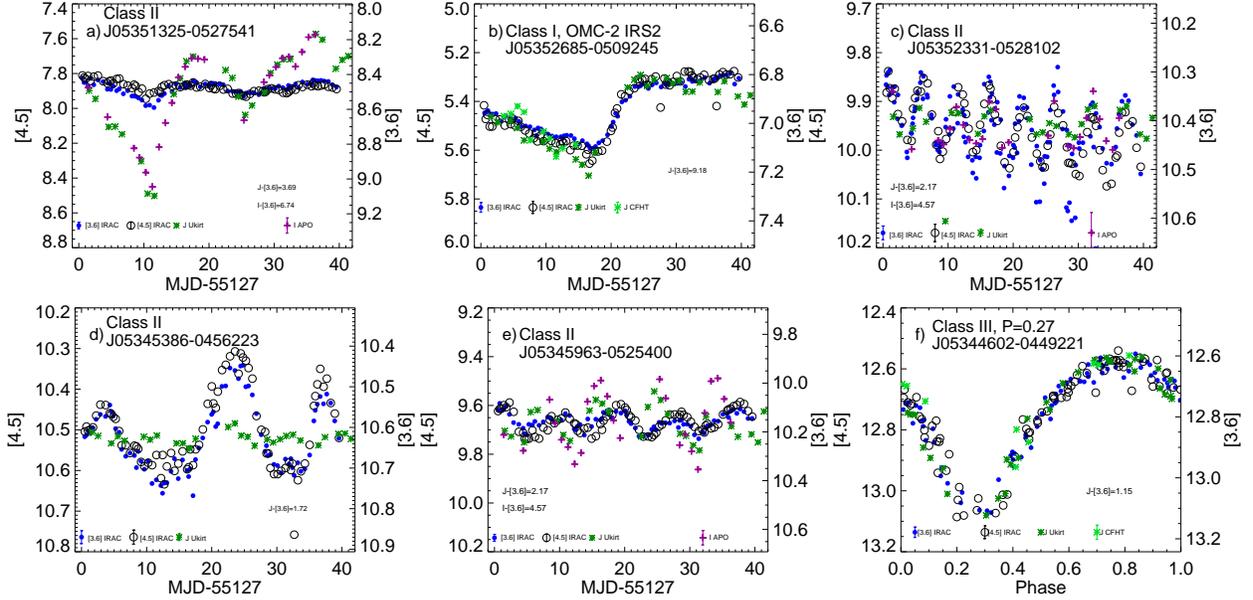}
\caption{Sample light curves for several stars showing shorter wavelength ground-based as well as Spitzer data selected to display the different variability types found in our data: a) larger amplitudes at shorter wavelengths; b) similar amplitudes of variation at all wavelengths; c) larger IRAC amplitudes than the shorter wavelength amplitudes; d) IRAC variations that are non-apparent at shorter wavelengths; e)difference in phase between the optical and near-IR data relative to IRAC data. Panel f) shows the phased light curve of star 1165 of \citet{Stassun99}. The symbols are $\bullet$: [3.6]; $\circ$: [4.5]; $*$: $J$ (dark green: UKIRT, light green: CFHT); and $+$: $I_c$ (pink: SLOTIS, purple: APO). [3.6] and [4.5] magnitudes have been plotted in the right and left vertical axis respectively. $I_c$ and $J$ light curves have been shifted in the y-axis to match the mean IRAC values. The amount shifted is stated in each panel.  \label{IJIRAClcs}}
\end{figure}

Fig.~\ref{IJIRAClcs} illustrates the value of having contemporaneous optical/NIR/Mid-IR monitoring (12\%, 85\%, and 37\% of the Orion stars have K$_s$, J and I$_c$ band light curves respectively). When compared to the near-IR and optical data, the mid-IR light curves usually have similar shapes. We have derived the amplitude of the light curves at each band as the difference between the maximum and minimum values of the magnitude at that bandpass avoiding single deviant datapoints. There are generally larger amplitudes at shorter wavelengths ($\sim$60\% of the variables with shorter wavelength data-- Fig.~\ref{IJIRAClcs}a), though in some cases, the amplitudes of variation are similar at all bands ($\sim$30\% of the variables with shorter wavelength data-- Fig.~\ref{IJIRAClcs}b). In a small number of cases, the IRAC variations are smaller or non-apparent at shorter wavelengths, indicating that the source of variability is probably in the disk ($\sim$10\% of the variables with shorter wavelength data-- Fig.~\ref{IJIRAClcs}cd). In only two cases do we find a phase difference between the optical and near-IR data relative to IRAC data (Fig.~\ref{IJIRAClcs}e). Some of these effects can be explained by the presence of one or more hot spots on the surface of the star and/or the existence of warps in its disk. 

Several of the objects deserve particular mention. Fig.~\ref{IRAClcs}d shows the well-known intermediate mass protostar OMC-2 IRS 1, for which outflows have been detected \citep{Yu97,Reipurth99,Takahashi08,Anathpindika08}; a similar light curve was previously reported for Orion BN/KL (J053514.12-052222.9 --\citet{Hillenbrand01}), though the physical mechanism for such variability in these B-type \citep{Johnson90,Hillenbrand01} YSOs is not obvious. In Fig.~\ref{IJIRAClcs}b, we show the protostar OMC-2 IRS2, which has also been associated with a jet \citep{Yu97,Anathpindika08} but for which only one scattering lobe is seen, suggesting that its disk is significantly less inclined to the line-of-sight \citep{Rayner89}. Fig.~\ref{IJIRAClcs}f shows the lightcurve of a star identified by \citet{Stassun99} as the fastest rotator in a study of the period distribution in the ONC (P=0.27 days, Star 1161 of \citet{Stassun99} (RA 05 34 45.93, Dec -04 49 22.0, from his Table 1). Given its spectral type (K5; \cite{Rebull01}), at 1-2 Myr, this period is shorter than break-up.   However, the $VI_cJHK$ photometry for the star suggest an age older than 30 Myr if it is at Orion distance. Because the light curve shape and amplitude in \citeauthor{Stassun99} appears identical to our 2009 light curve, we suspect this object is a tidally distorted binary, but it could also be a single near ZAMS rapid rotator (similar to HII1883 in the Pleiades).

In general, our variable YSOs show colorless changes at IRAC bands ($\sim$88\% of our sample show IRAC [3.6]$-$[4.5] RMS amplitudes smaller than 0.05 mag). The objects that do show IRAC color variations usually become redder as they fade. The typical peak-to-peak amplitudes in each IRAC band are $\sim$0.2 mag. The most extreme stars in our sample have changes in amplitude as large as $\sim$1.8 mag and the largest change in color is $\sim$0.3 mag. 

We have used the NStED Periodogram Service\footnotemark to derive periods for our sample. This service uses the Lomb-Scargle \citep{Scargle82}, Box-fitting Least Squares \citep{Kovacs02} and Plavchan \citep{Plavchan08periodogram} algorithms for computing periodograms from time series data. Given the cadence of our data, the periodograms are formally reliable for periods between 1 and 40 days however we have found 11 sources with periods shorter than a day. Those sources all have periods from the literature that match our determinations and the light curves are so well phased that we believe the periods are real. We have not retained any period longer than 25 days. In this way we can determine a period for 23\% of the variable YSOs with IR excess; 7\% of these are Class Is and 93\% are Class IIs. Among the non-excess sources, $\sim$44\% are variables and the predominant variablity is periodic, probably due to photospheric spots; 33\% of the WTTs have periods discernible from the Spitzer data. In the total ONC sample (1249 Spitzer excess sources plus 820 Orion members not known to have IR excess), we find 150 new periods (see Table\ref{tab:periods}).  About 30\% of our total sample had previously published periods. Of these, we recover the published period for 57\% of the WTTs and 47\% for the YSOs with IR excess. This difference is likely due to the thermal emission from dust at IRAC wavelengths significantly degrading the sensitivity to photospheric phenomena but also because the stars themselves may change \citep{Rebull01}. 

\footnotetext{http://nsted.ipac.caltech.edu/periodogram/cgi-bin/Periodogram/nph-simpleupload}

Finally, we have identified 9 eclipsing binaries in our sample. Four of them, 2MASS 05352184-0546085 \citep{Stassun06}, V1174 Ori \citep{Stassun04}, Par1802 \citep{Cargile08}, and  JW 380 \citep{Irwin07} are already known. A fifth one, Theta 1 Ori E, is a known 
spectroscopic binary \citep{Herbig06} which was flagged as potentially eclipsing, but no confirmatory photometry had been obtained until now.   We are working to obtain spectroscopic confirmation of the four new eclipsing binary candidates and will report on their status in a future paper. Their coordinates and tentative periods can be found in Table~\ref{binaries}.





\section{``Dipper" Objects: Clouds in the disk?}

\subsection{Identification and Characterization}

\begin{figure}
\epsscale{1.}
\plotone{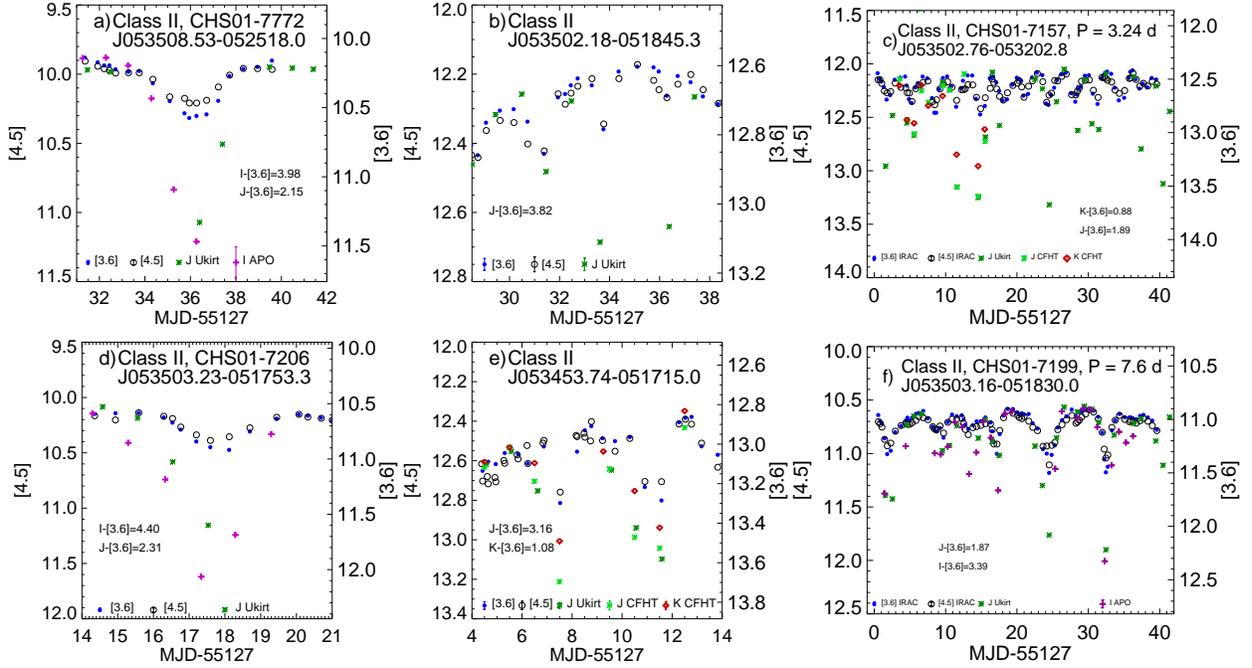}
\caption{Sample light curves for six of the stars showing flux dips that can be discerned against a steady or slowly varying continuum. The two leftmost objects show broad dips that are not repeated over the duration of the observation.  The two central panels show two of the narrowest dips and the two rightmost sources show periodic dips. The symbols are $\bullet$: [3.6]; $\circ$: [4.5]; $\diamond$: $K_s$; $*$: $J$ (dark green: UKIRT, light green: CFHT); and $+$: $I_c$ (pink: SLOTIS, purple: APO). [3.6] and [4.5] magnitudes have been plotted in the right and left vertical axis respectively. $I_c$, $J$, and $K_s$ light curves have been shifted in the y-axis to match the mean IRAC values. The amount shifted is stated in each panel.  \label{dippers}}
\end{figure}

One of the most surprising and interesting classes of variability we find is narrow flux dips. This subset of 41 variables shows brief, sharp drops in their flux density with durations of one to a few days. These objects are members of the ``Type III" variable class identified by \citet{Herbst94} based on an optical synoptic data set; their most famous prototype is AA Tau \citep{Bouvier03}.  A few examples of such ``dipper" light curves are shown in Figure~\ref{dippers}.   Among the identified cases, some exhibit only one dip, but typically $>$1 dip is seen over the 40 days.  Periodic dips are also detected in about a third of these objects, with periods from ~$\sim$2 to 14 days.  The dip amplitude is typically several tenths of a magnitude in the IRAC bands and ranges from 0.16-1.5 mag at $J$. 

The criteria imposed for selecting dipper stars were:  
1) a distinctive dip extends over several IRAC epochs unless a single epoch IRAC dip is corroborated by the independent but contemporaneous shorter wavelength data.  Implicitly, this means the IRAC-only dips will be broader in time, on average, than dips where shorter wavelength data confirm the dip.
2) If based on just an apparent dip in one IRAC epoch, both IRAC channels must agree 
and the I$_c$ or J-band data must show a larger amplitude dip by at least 50\% to avoid misidentification of eclipsing binary stars as AA Tau analogs.  
3) The ``continuum" must be flat enough so that the IRAC dip ``stands out." 
Although their identification by eye is generally straightforward, it is not as straightforward to determine the frequency of such dips in the overall data set.  To do so, one must account for the likely superposition of such events with other types of variability that are occurring in the same stars. The optical and near-IR data  were obtained at intervals of one to a few days as compared with the IRAC data which were obtained at half a day intervals, thus preventing confirmation of some IRAC dips.  To detect these events requires good S/N data in all the relevant bandpasses, which eliminates many of the fainter stars and/or stars in regions where the nebular background is bright or very structured. 

Based on the above criteria, we identify 38 of the Class I or II YSOs as having one or more AA Tau-like flux dips.  Given the coverage, cadence and S/N that we have in different filters, we estimate that we would be able to identify $\sim$70\% of broad dips (Fig.~\ref{dippers}ab) and about 40\% of the narrow dips (those present in just one IRAC epoch; Fig.~\ref{dippers}de). Thus, at present, we estimate the overall frequency of dipper stars at $\sim$5\% or higher. We note that this fraction is much lower than the 28\% estimated by \citet{Alencar10} for periodic ``AA Tau like" behavior in optical wavelength CoRoT data among a smaller set of young stars in NGC 2264. There are, in fact, a variety of reasons that CoRoT detects a larger fraction of "dipper" sources in NGC 2264 than the Spitzer YSOVAR survey of the ONC: CoRoT's much better relative photometric accuracy ($<$ 1 mmag) and higher cadence (every 512 sec during 23 days of uninterrupted observations), the larger amplitude of these dips at shorter wavelengths, the selection criteria itself, and the greater complexity of the mid-IR light curves due to the disk contribution. However, there may also be some real difference between the YSOs in the two clusters. Examination of the \citet{Carpenter01} monitoring NIR data reveals that $\sim$50\% of our dippers displayed similar behavior 10 years ago, suggesting some stability in physical mechanism inducing the variations.
We have identified 13 objects, or 34\% of the ``dipper" sample, as being periodic with P$<$40 days (the observing window). 
\citet{Carpenter01} derived similar periods (less than 0.6 days of difference and usually less than 0.1days) for five of these objects. The dependence of depth vs.\ wavelength is roughly consistent with that expected for the standard extinction law, as previously reported for AA Tau \citep{Bouvier03}. Coordinates, periods, and amplitudes for the dippers can be found in Table~\ref{dipperstable}.

Considering other existing data for our ONC stars, specifically the Spitzer photometry including all IRAC bands and MIPS 24 $\mu$m, 2MASS near-infrared, and literature optical photometry and spectral types, we find that there is no clear distinction of the AA Tau analogs in either Spitzer/IRAC or 2MASS magnitudes and colors relative to the rest of the ONC population, other than these objects are most often associated with Class II type YSOs.  Only one of the 129 Class Is vs.\ 37 of the 1175 Class IIs are identified as AA-Tau analogs. 

\subsection{Interpretation}

The AA Tau variability phenomenon has been interpreted \citep{Bertout00, Bouvier03} as arising from the rotation of a circumstellar disk with a high latitude ``warp" that periodically obscures the star in addition to the extinction due to a flared disk that is typically imposed over the rest of the orbit. However, any process which creates over-dense asymmetric regions in the inner disk could produce the flux dips.

Our interpretation is that these objects are being extincted by either ``clouds" of relatively higher opacity in the disk atmosphere or geometric warps in the inner disk.  In either case the dips are caused as the feature passes through our line of sight to the star as the disk rotates. This hypothesis is consistent with both the event duration and the variation of dip amplitude with wavelength. Further, as some of the dips are periodic or semi-periodic, we note that the derived periods are consistent with those expected from an inner disk region in co-rotation with a star having typical rotation rates for a Class II star (2-14 days). Similar features located further out in the disk could be responsible for the non-repeating and/or broader dips. 

As an example, for a period of P=8 days, the ``cloud", a region of enhanced dust column density, should be located at a Keplerian orbital radius 0.07 AU (or 7 R$_*$, for a typical 1 Myr M2 star; \citet{Baraffe98}) and for P$\geq$40 days (i.e., objects showing just one event), the occulting source would be located further than 0.19 AU  (or 18 R$_*$). We can roughly estimate the size and mass of the clouds by using the depth of the events, duration, and standard extinction law. Thus, for an object like that in Fig.~\ref{dippers}f, with P$\sim$8 days and dip duration $\sim$2 days, the size of the occulting body would be about 12 R$_*$. This large size is better associated with the warped disk model than with a discrete cloud in the disk. Given the average $J$ band depth of $\sim$ 1 mag, for the narrowest dips (i.e., dip duration $\sim$12 hr), the occulting body must have a radius $>$0.6 R$_*$. If we consider the cloud as a clump of dust grains, we could estimate its mass using the N(H) to $A_v$ relation from \citet{Gagne95}. If the cloud had a radius $\sim$1 R$_*$, its mass would be $\sim 2\times 10^{18}$ g. If the cloud were as large as 5 R$_*$, then the mass of the cloud increases to $\sim2.5\times 10^{20}$ g. 
These masses are small
relative to the Moon (Mass$_{Moon}$=7.35$\times10^{25}$g), hence, these clouds are probably not the source material
to form a planet.

A problem with the disk obscuration interpretation arises if the flux dips are seen in diskless WTTs.  Indeed, there are 3 ONC members not previously known to have IR excesses (SOY 053524.15-044930.1, SOY 053529.02-050604.0, and SOY 053516.74-052019.7; see Table~\ref{dipperstable}) which have apparent AA-Tau type flux dips. The two first stars show just one apparent flux dip; the third shows 5 dips with no clear period. Examination of the SEDs of these stars shows that the latter shows evidence of having an IR excess at 5.8 and 8~$\mu$m (though not at 3.6 and 4.5~$\mu$m), consistent with it having a disk with a large inner hole.  In this scenario, if the inner disk edge is sufficiently far from the star, there could be no excess at $<$5~$\mu$m, but there still could be material which could occult our view of the star and  cause a flux dip.  The other two stars could have even larger inner disk holes.  

\section{Conclusions}

The YSOVAR Orion data provides the first large-scale survey of the photometric variability of YSOs in the mid-IR. In Fall 2009, Spitzer/IRAC observed at 3.6 and 4.5 $\mu$m a 0.9 sq. deg. area centered on the Trapezium cluster twice a day for 40 consecutive days, producing light curves for 1249 known disked YSOs and $\sim820$ young but generally diskless WTTs, with typical accuracies $\sim$3\%. These data are publicly accessible through the YSOVAR website (http://ysovar.ipac.caltech.edu/) or in Table~\ref{TimeSeries}.

We find that 70\% of the disked YSOs and 44\% of the WTTs are variable. The WTT variations are mostly spot-like, while there is a wide range of variability types for the YSOs. Accordingly, 79\% of the variable WTTs are periodic, while periods are detected for only 24\% of the YSOs.  On Table~\ref{tab:periods} we list 150 new periods found based on these data. We identify five new PMS eclipsing binary candidates, including one already known spectroscopic binary, Theta Ori E, and another one that is fainter than all previously known ONC PMS eclipsing binaries. Coordinates and preliminary periods for the four new eclipsing binary candidates are given in Table~\ref{binaries}.

We have discussed in detail a sample of the Orion variables that show short duration flux dips (AA Tau events). The mean magnitudes of these stars plus the depth of the dips and periods are given in Table~\ref{dipperstable}. We interpret these events as the star being extincted by either clouds of relatively higher opacity in the disk atmosphere or geometric disk warps of relatively higher latitude which pass through the line of sight of the star. This hypothesis is consistent with both the  duration and the wavelength dependent behavior of the events.




\acknowledgments
We would like to thank the anonymous referee for her helpful comments.
This work is based on observations made with the Spitzer Space Telescope, which
is operated by the Jet Propulsion Laboratory, California Institute of Technology under a contract
with NASA. This work is also based [in part] on observations obtained with WIRCam, a joint project of CFHT, Taiwan, Korea, Canada, France, and the Canada-France-Hawaii Telescope (CFHT) which is operated by the National Research Council (NRC) of Canada, the Institute National des Sciences de l'Univers of the Centre National de la Recherche Scientifique of France, and the University of Hawaii.



{\it Facilities:} \facility{Spitzer (IRAC)}

\clearpage



\clearpage

\clearpage

\begin{deluxetable}{ccccccc}
\tabletypesize{\scriptsize}
\rotate
\tablecaption{Summary of observations\label{Data}}
\tablewidth{0pt}
\tablehead{
\colhead{Telescope/Instrument} & \colhead{Filter} & \colhead{Start/End dates} & \colhead{N epochs} &\colhead{Central coordinates (J2000)}  &\colhead{Area (arcmin)}  & \colhead{Exptime/epoch (sec.)}
}
\startdata
Spitzer/IRAC        & &Oct. 23 - Dec. 1, 2009                & 81  &05:35:10 -05:42:25& 30$\times$35       &(10.4\&0.4)$\times$4\\
Spitzer/IRAC        && Oct. 23 - Dec. 1, 2009               &81  &05:36:06 -05:16:22& 5$\times$25        &(10.4\&0.4)$\times$4\\
Spitzer/IRAC        &[3.6],[4.5] & Oct. 23 - Dec. 1, 2009              & 81 &05:35:21 -04:49:47&  30$\times$35      &(10.4\&0.4)$\times$4\\
Spitzer/IRAC        &&Oct. 23 - Dec. 1, 2009               &  81 &05:34:37 -05:20:00&   10$\times$25     &(10.4\&0.4)$\times$4\\
Spitzer/IRAC        &&Oct. 23 - Dec. 1, 2009               & 81  &05:35:24 -05:17:32&   20$\times$25     &1.2$\times$20\\
\hline
UKIRT/WFCAM       & $J$          & Oct. 20 - Dec. 22, 2009           & 32 &05:35:17 -05:22:49& 52$\times$104    & 2$\times$3\\
\hline
CFHT/WIRCam   &  &Oct. 27 - Nov. 8, 2009             &11&05:35:14 -04:41:34& &\\
CFHT/WIRCam   &   &Oct. 27 - Nov. 8, 2009             &11 &05:35:14 -05:02:14& &\\
CFHT/WIRCam   & $J,Ks$    & Oct. 27 - Nov. 8, 2009              & 11 &05:36:04 -05:22:46&21$\times$21 &(5$\times$2)$\times$7(J), 5$\times$7(K$_s$)\\
CFHT/WIRCam   &    & Oct. 27 - Nov. 8, 2009        &11 &05:34:26 -05:22:46& &\\
CFHT/WIRCam   &   & Oct. 27 - Nov. 8, 2009        &11 &05:35:14 -05:43:28& &\\
\hline
NMSU APO1m/CCD       & $I_c$           & Oct. 24, 2009 - Feb. 09, 2010 &27 &05:35:18 -05:24:03&15$\times$15 &60$\times$20\\
\hline
SuperLotis/CCD &$I_c$            & Oct. 28 - Dec. 12, 2009            & 23&05:35:11 -05:41:02 &16$\times$16 &180$\times$12\\
\enddata

\label{data}
\end{deluxetable}

\clearpage
\begin{deluxetable}{lcccccc}
\tabletypesize{\scriptsize}
\tablecolumns{6}
\tablewidth{0pt}
\tablecaption{Time series at the [3.6] and [4.5] IRAC bands for the ONC stars. (Full table available in the electronic version of this article.) \label{TimeSeries}}
\tablehead{ 
\colhead{Object} &\colhead{MJD (days)} &\colhead{IRAC band} &\colhead{mag} &\colhead{ error} &\colhead{RA J2000 (deg)} &\colhead{Dec. J2000 (deg)}}
\startdata
 ISOY\_J053532.71-045011.9 & 55127.50781250   &IRAC1      &8.725      &0.006   &83.886268   &-4.836648\\
 ISOY\_J053532.71-045011.9  &55127.76562500   &IRAC1      &8.715      &0.006   &83.886268   &-4.836648\\
 ISOY\_J053532.71-045011.9  &55128.00781250   &IRAC1      &8.704      &0.006   &83.886268   &-4.836648\\
 ISOY\_J053532.71-045011.9  &55128.37890625   &IRAC1      &8.741      &0.004   &83.886268   &-4.836648\\
 ISOY\_J053532.71-045011.9  &55128.76171875   &IRAC1      &8.814      &0.006   &83.886268   &-4.836648\\
 ISOY\_J053532.71-045011.9  &55129.25781250   &IRAC1      &8.795      &0.005   &83.886268   &-4.836648\\
 ISOY\_J053532.71-045011.9  &55129.89843750   &IRAC1      &8.823      &0.006   &83.886268   &-4.836648\\
 ISOY\_J053532.71-045011.9  &55130.69921875   &IRAC1      &8.905      &0.005   &83.886268   &-4.836648\\
 ISOY\_J053532.71-045011.9  &55131.35937500   &IRAC1      &8.925      &0.007   &83.886268   &-4.836648\\
 ISOY\_J053532.71-045011.9  &55131.58593750   &IRAC1      &8.912      &0.006   &83.886268   &-4.836648\\
 \enddata

\end{deluxetable}

\clearpage

\begin{deluxetable}{rccc}
\tabletypesize{\scriptsize}
\tablecaption{Statistics}
\tablewidth{0pt}
\tablehead{
\colhead{} & \colhead{Class I} & \colhead{Class II} & \colhead{Class III}} 
\startdata
\# sources & 126 & 1123 & 820\\
\hline
IRAC Variables & 106 & 787 & 366\\
Variables w/ [3.6] amp. $>$0.2 mag &71 & 465&92 \\
Variables w/ [3.6] amp. $>$0.5 mag &15 & 59&8 \\
Variables w/ [3.6] amp. $>$1 mag & 3     & 1& 0\\
$[3.6]-[4.5]$ Color Variables w/ amp. $>$0.2 mag& 3 &5 &0 \\
\hline
Prev. known Periods & 8& 266&334 \\
Recovered Periods & 5& 125& 189\\
New Periods & 8& 76& 98\\
\hline
\# AA Tau analogs & 1& 37& 3\\
\enddata
\label{statistics}
\end{deluxetable}

\begin{deluxetable}{rcc}
\tabletypesize{\scriptsize}
\tablecaption{New periods for Orion members}
\tablewidth{0pt}
\tablehead{
\colhead{Source\tablenotemark{a}} & \colhead{Period (days)}& \colhead{p-value}\tablenotemark{b} } 
\startdata
  ISOY\_J053427.70-053155.4 & 3.37 & 4.78E-4\\
  ISOY\_J053434.88-044243.5 & 9.36 & 1.31E-4\\
  ISOY\_J053435.15-053210.4 & 5.18 & 2.46E-5\\
  ISOY\_J053435.98-045218.0 & 2.34 & 2.80E-3\\
  ISOY\_J053439.11-053839.2 & 2.41 & 3.01E-2\\
  ISOY\_J053439.76-052425.6 &16.97 & 1.65E-4\\
  ISOY\_J053441.18-054611.7 & 8.02 & 1.18E-3\\
  ISOY\_J053441.71-052653.0 & 8.47 & 9.13E-2\\
  ISOY\_J053441.87-055502.6 & 1.66 & 3.23E-6\\
  ISOY\_J053442.74-052837.6 & 6.15 & 1.40E-7\\
  ISOY\_J053445.00-045559.2 & 3.40 & 3.60E-3\\
  ISOY\_J053447.64-054350.7 & 4.27 & 1.35E-2\\
  ISOY\_J053447.74-052632.1 & 1.84 & 5.73E-6\\
  ISOY\_J053447.85-044228.9 & 8.36 & 8.55E-4\\
  ISOY\_J053449.57-052903.4 &14.73 & 9.90E-7\\
  ISOY\_J053450.66-050407.7 & 9.24 & 2.03E-3\\
  ISOY\_J053450.72-045836.8 & 6.32 & 3.80E-7\\
  ISOY\_J053452.22-045102.7 & 5.15 & 1.52E-2\\
  ISOY\_J053452.49-044940.4 & 6.25 & 5.78E-2\\
  ISOY\_J053452.60-051536.6 & 2.35 & 3.00E-7\\
  ISOY\_J053453.12-055955.0 & 2.55 & 3.09E-2\\
  ISOY\_J053453.99-044537.3 & 2.38 & 1.00E-8\\
  ISOY\_J053454.83-052512.6 & 3.75 & 2.88E-4\\
  ISOY\_J053454.90-054643.9 &13.55 & 4.70E-7\\
  ISOY\_J053454.92-050128.4 &13.90 & 2.89E-2\\
  ISOY\_J053457.20-050823.9 & 1.63 & 2.85E-5\\
  ISOY\_J053458.53-060000.4 & 5.98 & 6.03E-2\\
  ISOY\_J053500.22-052409.2 & 7.29 & 2.60E-4\\
  ISOY\_J053500.54-054859.1 & 6.63 & 9.01E-6\\
  ISOY\_J053501.08-052304.1 & 3.98 & 6.39E-5\\
  ISOY\_J053501.10-052337.3 & 7.12 & 4.55E-3\\
  ISOY\_J053501.34-052811.9 & 7.23 & 7.00E-8\\
  ISOY\_J053501.34-054113.5 & 8.41 & 1.36E-5\\
  ISOY\_J053502.38-051548.0 & 6.09 & 1.49E-6\\
  ISOY\_J053503.12-050917.0 & 4.32 & 5.17E-5\\
  ISOY\_J053503.24-051726.4 & 4.73 & 5.13E-6\\
  ISOY\_J053503.70-052245.7 & 5.11 & 1.00E-8\\
  ISOY\_J053504.56-052013.9 & 8.43 & 1.40E-7\\
  ISOY\_J053504.61-045829.0 & 4.67 & 1.04E-3\\
  ISOY\_J053504.80-060020.8 & 5.13 & 1.83E-3\\
  ISOY\_J053504.95-052109.2 &10.12 & 2.00E-8\\
  ISOY\_J053505.45-052230.5 & 3.56 & 2.25E-4\\
  ISOY\_J053505.71-052354.1 &20.48 & 5.41E-1\\
  ISOY\_J053505.79-053827.8 & 2.80 & 3.50E-7\\
  ISOY\_J053505.85-044843.3 & 6.16 & 1.63E-2\\
  ISOY\_J053506.77-055101.3 & 6.76 & 1.50E-4\\
  ISOY\_J053506.86-051133.2 & 5.47 & 1.04E-6\\
  ISOY\_J053507.01-050134.8 & 2.83 & 8.27E-3\\
  ISOY\_J053507.70-055749.5 & 3.75 & 3.89E-2\\
  ISOY\_J053508.02-053244.3 & 6.87 & 2.13E-4\\
  ISOY\_J053508.07-054853.8 & 7.97 & 2.78E-4\\
  ISOY\_J053508.26-055000.3 & 6.83 & 1.04E-6\\
  ISOY\_J053509.95-051449.9 & 2.19 & 8.94E-6\\
  ISOY\_J053510.02-051816.6 & 9.37 & 1.00E-8\\
  ISOY\_J053510.99-051521.8 &11.29 & 3.53E-3\\
  ISOY\_J053511.10-051601.8 & 4.75 & 2.00E-8\\
  ISOY\_J053511.48-052352.0 &22.37 & 1.49E-2\\
  ISOY\_J053511.56-052448.0 & 3.74 & 7.68E-2\\
  ISOY\_J053511.62-051912.3 & 6.77 & 2.36E-2\\
  ISOY\_J053511.89-052002.0 & 4.66 & 1.40E-6\\
  ISOY\_J053511.95-052845.0 & 3.11 & 1.13E-4\\
  ISOY\_J053513.46-053502.8 & 7.28 & 9.36E-4\\
  ISOY\_J053513.57-053508.1 & 5.73 & 3.24E-2\\
  ISOY\_J053514.66-050312.6 & 7.66 & 2.28E-4\\
  ISOY\_J053514.88-055142.2 & 4.27 & 6.33E-4\\
  ISOY\_J053515.48-052722.7 & 6.54 & 2.84E-6\\
  ISOY\_J053515.49-053511.9 & 5.76 & 3.71E-2\\
  ISOY\_J053515.52-055316.2 &10.43 & 6.00E-8\\
  ISOY\_J053515.56-045308.1 &13.22 & 3.90E-7\\
  ISOY\_J053515.65-055214.6 & 8.67 & 3.49E-4\\
  ISOY\_J053515.76-052309.9 & 9.90 & 3.00E-8\\
  ISOY\_J053516.33-052932.7 &16.79 & 3.08E-6\\
  ISOY\_J053516.98-053246.4 & 3.77 & 1.42E-2\\
  ISOY\_J053517.10-051900.8 &17.46 & 2.17E-3\\
  ISOY\_J053517.21-044113.6 & 1.05 & 6.44E-2\\
  ISOY\_J053517.35-052235.8 & 7.10 & 2.35E-2\\
  ISOY\_J053517.40-045957.2 & 3.00 & 9.60E-2\\
  ISOY\_J053517.53-051929.0 & 6.72 & 2.48E-5\\
  ISOY\_J053517.54-051613.1 & 4.06 & 7.10E-7\\
  ISOY\_J053517.97-054937.5 & 6.27 & 5.94E-2\\
  ISOY\_J053518.03-052205.4 & 5.63 & 1.06E-5\\
  ISOY\_J053518.23-051745.0 & 3.28 & 4.39E-6\\
  ISOY\_J053518.29-050805.0 & 4.85 & 2.71E-4\\
  ISOY\_J053518.70-051801.8 & 7.37 & 1.97E-4\\
  ISOY\_J053519.86-053103.8 & 4.23 & 6.85E-6\\
  ISOY\_J053520.02-052911.9 & 2.79 & 1.37E-2\\
  ISOY\_J053520.06-050358.5 &12.41 & 9.96E-4\\
  ISOY\_J053520.19-052308.6 & 9.02 & 3.32E-3\\
  ISOY\_J053520.29-054639.9 & 7.90 & 5.73E-2\\
  ISOY\_J053520.72-051926.5 &22.21 & 1.00E-8\\
  ISOY\_J053521.38-050942.3 & 3.76 & 9.00E-8\\
  ISOY\_J053521.62-052325.8 &13.72 & 3.39E-5\\
  ISOY\_J053522.10-051857.6 & 5.78 & 8.76E-6\\
  ISOY\_J053522.32-044133.3 &10.29 & 9.60E-2\\
  ISOY\_J053522.73-051838.2 & 5.75 & 5.20E-7\\
  ISOY\_J053522.98-051521.8 & 4.32 & 1.33E-3\\
  ISOY\_J053523.04-052941.5 & 2.94 & 1.30E-7\\
  ISOY\_J053523.18-052228.3 & 8.33 & 5.18E-3\\
  ISOY\_J053523.97-055942.0 &18.70 & 7.13E-5\\
  ISOY\_J053524.33-050120.5 &14.65 & 1.23E-4\\
  ISOY\_J053524.34-052232.3 &18.91 & 3.02E-6\\
  ISOY\_J053524.62-052104.2 &16.78 & 7.10E-3\\
  ISOY\_J053524.67-044943.0 &17.15 & 9.00E-7\\
  ISOY\_J053524.73-051030.1 & 4.37 & 3.47E-4\\
  ISOY\_J053525.18-050509.3 & 5.64 & 9.42E-5\\
  ISOY\_J053526.06-052121.0 & 2.58 & 1.28E-4\\
  ISOY\_J053526.47-053016.4 & 3.70 & 4.34E-5\\
  ISOY\_J053526.88-044730.7 & 3.91 & 3.07E-5\\
  ISOY\_J053527.00-054845.8 &11.27 & 4.94E-2\\
  ISOY\_J053527.27-050527.0 &17.86 & 6.35E-3\\
  ISOY\_J053527.66-054255.1 &12.31 & 6.22E-4\\
  ISOY\_J053528.19-050341.2 &23.52 & 3.48E-5\\
  ISOY\_J053529.65-052002.2 & 2.44 & 4.66E-6\\
  ISOY\_J053530.61-045936.1 & 8.57 & 3.05E-4\\
  ISOY\_J053532.67-054528.4 & 2.53 & 6.30E-7\\
  ISOY\_J053532.71-045011.9 &14.38 & 4.32E-6\\
  ISOY\_J053532.96-051204.7 & 1.30 & 1.35E-4\\
  ISOY\_J053533.34-051145.6 &16.19 & 1.40E-7\\
  ISOY\_J053533.60-052209.8 &21.48 & 1.20E-7\\
  ISOY\_J053534.03-055505.8 &11.47 & 1.04E-4\\
  ISOY\_J053535.16-051946.9 & 4.12 & 1.12E-2\\
  ISOY\_J053535.22-044739.6 &10.24 & 3.88E-2\\
  ISOY\_J053536.49-044259.2 & 2.46 & 2.34E-2\\
  ISOY\_J053536.50-052009.4 & 9.87 & 3.97E-4\\
  ISOY\_J053538.56-050803.4 &16.77 & 3.49E-6\\
  ISOY\_J053539.07-052025.7 & 4.51 & 1.82E-2\\
  ISOY\_J053539.49-044019.3 & 3.31 & 3.53E-3\\
  ISOY\_J053539.74-045141.7 & 3.05 & 6.27E-7\\
  ISOY\_J053539.77-044024.3 &17.91 & 6.18E-3\\
  ISOY\_J053540.15-053723.7 & 6.31 & 4.05E-2\\
  ISOY\_J053541.70-052014.9 & 6.67 & 1.72E-2\\
  ISOY\_J053541.79-045027.7 &24.80 & 1.26E-3\\
  ISOY\_J053542.01-051011.5 & 8.21 & 6.91E-4\\
  ISOY\_J053544.09-050837.6 & 4.46 & 1.40E-7\\
  ISOY\_J053545.32-044334.9 & 2.36 & 4.33E-2\\
  ISOY\_J053546.23-051808.6 & 9.71 & 2.00E-8\\
  ISOY\_J053547.80-051030.8 & 2.26 & 2.47E-6\\
  ISOY\_J053551.07-051508.9 & 9.36 & 4.56E-5\\
  ISOY\_J053552.10-044915.1 & 3.76 & 9.46E-4\\
  ISOY\_J053553.21-044734.7 & 1.73 & 2.99E-2\\
  ISOY\_J053554.63-052707.5 & 8.45 & 6.27E-7\\
  ISOY\_J053556.10-052533.0 & 5.17 & 2.46E-5\\
  ISOY\_J053558.94-053253.9 & 8.31 & 8.38E-4\\
  ISOY\_J053559.17-053552.7 &19.04 & 1.54E-3\\
  ISOY\_J053605.95-050041.2 & 3.56 & 6,04E-3\\
  ISOY\_J053606.67-045512.1 & 1.66 & 1.01E-4\\
  ISOY\_J053606.99-051334.2 & 2.65 & 5.66E-6\\
  ISOY\_J053609.28-045000.8 &15.16 & 1.07E-2\\
  ISOY\_J053612.32-045819.0 & 4.09 & 9.61E-2\\
  ISOY\_J053614.94-051533.7 & 4.72 & 4.05E-3\\
\hline
\enddata
\label{tab:periods}
\tablenotetext{a}{ISOY stands for Initial Spitzer Orion YSO.}
\tablenotetext{b}{The p-value is the probability of chance occurrence returned by the periodogram service.}
\end{deluxetable}

\begin{deluxetable}{rc}
\tabletypesize{\scriptsize}
\tablecaption{New PMS Eclipsing Binary candidates}
\tablewidth{0pt}
\tablehead{
\colhead{Source\tablenotemark{a}} & \colhead{Period (days)}} 
\startdata
ISOY\_J053526.88-044730.7 & 3.91\\
ISOY\_J053518.03-052205.4 & 5.63\\
ISOY\_J053505.71-052354.1 & 20.39\\
ISOY\_J053605.95-050041.2 & 3.56\\
\hline
\enddata
\label{binaries}
\tablenotetext{a}{ISOY stands for Initial Spitzer Orion YSO.}
\end{deluxetable}

\begin{deluxetable}{rccccccccc}
\rotate
\tabletypesize{\scriptsize}
\tablecaption{ONC AA Tau analogs}
\tablecolumns{10} 
\tablewidth{0pc}
\tablehead{
\colhead{Source\tablenotemark{a}} & \colhead{mean I}  & \colhead{mean J} & \colhead{mean [3.6] } & \colhead{mean [4.5]} & \colhead{Dip depth I\tablenotemark{b}} & \colhead{Dip depth J\tablenotemark{b}} &  \colhead{Dip depth [3.6 ]\tablenotemark{b}} & \colhead{Dip depth [4.5]\tablenotemark{b}}  & \colhead{Dip Period} \\\colhead{ } &\colhead{mag} & \colhead{mag} & \colhead{mag} & \colhead{mag} & \colhead{mag} & \colhead{mag} & \colhead{mag} & \colhead{mag}&\colhead{days}}
\startdata
ISOY\_053432.02-052742.7 &    --   &13.255 & 10.437 & 10.015 &---  & 0.7 &0.2   & 0.17   &  --\\
ISOY\_053435.98-045218.0 &    --   &13.972 & 12.131 & 11.754 &---  & 0.6 & 0.2  & 0.15   &  2.34$\pm$0.12\\
ISOY\_053440.64-050658.7 &    --   &13.989 & 10.509 & 10.136 &---  & 1.1 & 0.5  &0.38    &  3.85$\pm$0.39\\
ISOY\_053446.79-052129.2 &    --   &13.424 & 11.213 & 10.862 &---  &0.42  &0.25   &0.20    &  --\\
ISOY\_053450.72-045836.8 &    --   &12.147 & 10.193 & 9.968  &---  &  0.34& 0.20  &0.26    &  6.32$\pm$0.38\\
ISOY\_053450.87-053929.2 &  14.319 &12.623 & 10.688 & 10.35  &---  &0.20  & 0.12  & 0.10   &  --\\
ISOY\_053453.74-051715.0 &    --   &16.316 & 13.158 & 12.675 &---  & 0.7 & 0.45  &  0.34  &  --\\
ISOY\_053455.10-044941.5 &    --   &13.253 & 11.09  & 10.749 &---  &0.85  & 0.27  & 0.18   & --\\
ISOY\_053502.18-051845.3 &    --   &16.589 & 12.772 & 12.347 &---  &  0.45& 0.16  & 0.13   &  --\\
ISOY\_053502.76-053202.8 &    --   &14.451 & 12.566 & 12.217 &---  &1.1  &0.35   & 0.25   &  3.24$\pm$0.78\\
ISOY\_053503.16-051830.0 &  14.424 &12.904 & 11.038 & 10.736 & 1.35 & 1.25 & 0.58  &0.45    &  7.57$\pm$0.72\\
ISOY\_053503.23-051753.3 &  15.098 &13.002 & 10.696 & 10.252 &1.40  & 1.15 & 0.35  &0.25    &  --\\
ISOY\_053503.96-051859.8 &  15.105 &13.069 & 10.841 & 10.423 & 0.95 &0.60  & 0.40  &  0.35  &  --\\
ISOY\_053504.68-044621.9 &    --   &15.488 & 11.697 & 11.148 &---  & 1.55 &  0.6 & 0.5   &  --\\
ISOY\_053504.86-054426.7 &  18.038 &15.385 & 11.701 & 11.305 & --- & 0.6 &0.5   &  0.4  &  --\\
ISOY\_053504.87-052057.5 &  16.334 &13.455 & 10.665 & 10.271 & 0.95 & 0.65 & 0.22  &    &  8.22$\pm$1.23\\
ISOY\_053504.90-053949.7 &    --   &15.952 & 13.007 & 12.658 &---  & 0.37 & 0.25  & 0.2   &  --\\
ISOY\_053507.53-051114.4 &    --   &12.409 & 9.991  & 9.783  &---  &0.32  &0.16   &   0.13 &  --\\
ISOY\_053508.07-054853.8 &  15.701 &13.16  & 9.927  & 9.45   &  1.25& 1.2 & 0.18  & 0.14   &  7.97$\pm$1.59\\
ISOY\_053508.53-052518.0 &  14.210 &12.376 & 10.231 & 9.973  & 1.3 & 1.07 & 0.40  & 0.28   &  --\\
ISOY\_053508.60-052619.4 &  11.956 &13.274 & 11.03  & 10.68  &0.25  &0.18  &0.1   & 0.1   &  --\\
ISOY\_053510.07-051706.8 &  13.722 &14.587 & 10.752 & 10.068 &1.1  &0.8  & 0.25  & 0.2   &  --\\
ISOY\_053510.19-052021.0 &  11.58  &13.571 & 10.938 & 10.642 & 0.6 & 0.45 &0.15   & 0.11   &  --\\
ISOY\_053510.94-043957.6 &    --   &13.636 & 11.342 & 11.075 &---  &1.25  & 0.3  & 0.3   &  --\\
ISOY\_053510.99-051521.8 &    --   &16.255 & 9.17   & 8.561  & --- &1.7  &0.6   &0.35    &  11.29$\pm$0.56\\
ISOY\_053516.33-051538.0 &    --   &11.961 & 9.617  & 9.354  & --- & 0.6 &0.23   &0.20    &  --\\
ISOY\_053516.74-052020.0 &    -- &13.275 & 9.621  & 9.34   &---  &0.8  &0.2   &0.18    &  --\\
ISOY\_053518.42-044000.2 &    --   &13.522 & 12.018 & 11.67  & --- &  0.28& 0.12  &---    &  --\\
ISOY\_053518.70-051801.8 &  12.457 &13.932 & 10.153 & 9.85   & 0.7 & 1.1 & 0.4  & 0.4   &  7.37$\pm$0.80\\
ISOY\_053523.98-052509.8 &  10.991 &13.227 & 10.6   & 9.774  & 0.35 & 0.35 & 0.17  & 0.11   &  --\\
ISOY\_053524.14-044930.3 &    --   &13.941 & 11.904 & 11.515 & --- &0.79  & 0.46  & 0.44   &  --\\
ISOY\_053525.51-054544.8 &  14.306 &12.887 & 9.816  & 9.38   & 1.6 & 1.25 & 0.30  &  0.25  &  7.57$\pm$0.31\\
ISOY\_053526.73-051645.1 &  11.22  &13.185 & 11.006 & 10.806 &1.1  & 0.6 &  0.26 & 0.24   &  2.96$\pm$0.62\\
ISOY\_053529.03-050604.1 &    --         &11.675 & 10.797 & 10.802 & --- & 0.24 & 0.06   &  0.06  &  --\\
ISOY\_053530.45-052811.3 &  12.139 &13.72 & 11.575 & 11.179 &   0.46   & 0.22 & 0.12  &0.10    &  --\\
ISOY\_053534.41-051838.6 &    --   &15.817 & 11.332 & 10.781 & --- &1.15  & 0.50  & 0.37   &  4.75$\pm$0.33\\
ISOY\_053535.61-053908.1 &  15.154 &13.793 & 11.142 & 10.653 &1.25  &1.15  &  0.6 & 0.5   &  --\\
ISOY\_053536.42-050115.5 &    --   &10.665 & 8.632  & 8.355  & --- &0.69  &  0.15 & 0.14   &  --\\
ISOY\_053547.65-053738.8 &    --   &12.808 & 10.413 & 10.068 & --- & 0.95 & 0.3  &  0.25  &  --\\
ISOY\_053556.10-052533.0 &    --   &15.613 & 12.045 & 11.727 & --- & 0.8 & 0.35  & 0.26   &  5.17$\pm$0.15\\
ISOY\_053558.94-053253.9 &    --   &14.063 & 11.466 & 10.99  & --- &1.05  &0.45   &0.42    &  --\\
\hline
\enddata
\label{dipperstable}
\tablenotetext{a}{ISOY stands for Initial Spitzer Orion YSO.}
\tablenotetext{b}{Dip depths are mean values for sources with more than one dip.}
\end{deluxetable}

\end{document}